\documentclass[12pt]{amsart}

\newtheorem{Theorem}{Theorem}
\newtheorem{Corollary}{Corollary}

\newtheorem{Proposition}{Propositon}
\newtheorem{Lemma}{Lemma}
\newcommand{\R}{\ensuremath{\mathbb{R}}}
\newcommand{\N}{\ensuremath{\mathbb{N}}}
\newcommand{\Z}{\ensuremath{{\mathsf{Z\!\!Z}}}}

\newcommand{\nn}{\{1, 2, \cdots , n\}}

\newcommand{\ep}{E(\pi)}

\begin{document}
  \title[Schr\"odinger Operators along  IET] {Cantor Singular Continuous Spectrum for Operators Along  Interval Exchange Transformations}
  \author{M. Cobo}
\address{ Departamento de Matem\'atica, UFES, Av.\ F.\ Ferrari 514,
Vit\'oria, ES, 19075-910 Brazil}
\email{miltonc@cce.ufes.br}
  \author{C. Gutierrez}
\address{ Departamento de Matem\'atica, ICMC/USP, CxP 668, S\~ao
Carlos,
SP, 13560-970 Brazil }
\email{gutp@icmc.usp.br}

\author{C. R. de Oliveira}
\address{Departamento de Matem\'{a}tica, UFSCar, S\~{a}o Carlos,  SP,
13560-970 Brazil}
\email{oliveira@dm.ufscar.br}

\begin{abstract} It is shown that Schr\"odinger operators, with
potentials along the shift embedding of
Lebesgue almost every interval exchange transformations, have Cantor
spectrum  of measure zero and
pure singular continuous for Le\-bes\-gue almost all points of the
interval.

\thanks{
}
\end{abstract}

\maketitle

\section{Introduction and Main Result}  In \cite{deOG} the  spectrum
$\sigma(H_\omega)$ of the discrete
Schr\"odinger operators
\begin{equation}\label{hamiltonian}
\begin{array}{ll}
&H_\omega:l^2(\Z)\to l^2(\Z) \\
& \\
&({H_{\omega}}\psi)_j := \psi_{j+1}+\psi_{j-1}+{\omega}_{j}\psi_j,
\end{array}
\end{equation}
 with $\omega=(\omega_j)_{j\in\Z}$ a  sequence  of real numbers (the
so-called potential) modulated along
the shift embedding of interval  exchange transformations (iets)
\cite{Keane1,KH} was investigated. There it was proved the  presence of
pure singular continuous spectrum
for $H_\omega$, for the shift associated with a  dense set of iets, where the potential
$\omega$ corresponds to the {\it itinerary} of the orbit of a.e.~$x$
in the  interval (from here on, a.e.\ with no specification means {\it almost everywhere}
with   respect to Lebesgue measure).

In this work we give a  step further by showing that the above
mentioned results of \cite{deOG} hold for Lebesgue almost
every iets. The proof of absence of eigenvalues involves a rather
different argument, mainly the Rauzy
induction map.  As a by-product of our a.e.\ results, which are
summarized
in Theorem~\ref{mainThm} ahead,
there is a set of interesting results in the literature ready to be
applied, which leads to Cantor
spectrum of zero Lebesgue measure; see Corollary~\ref{corolZero}.  \\

\noindent{\bf Interval exchanges.}
Let us recall some notations and a
description of  the~iets necessary to state and prove our results.

Given a semi-open interval $[a,b)$ and a vector
$\lambda=(\lambda_1,\lambda_2,\dots,\lambda_n)$ in $\R^n_+$ (i.e, all entries are positive)
such that $\lambda_1+\lambda_2+\dots+\lambda_n= b-a,$ we consider the
partition
of $[a,b)$ in the $n$ consecutive semi-open intervals
\[ I_1:=[a,a_1), I_2:=[a_1,a_2),\dots, I_n:=[a_{n-1},b) \] whose
lengths are respectively  $\lambda_1,\lambda_2,\dots,\lambda_n.$
Let $\pi$ be a permutation of the symbols $\{1,2,\dots,n\}.$ An iet
$E\colon [a,b)\to [a,b)$
is associated to the pair
$(\pi, \lambda)$ by exchanging the positions of the intervals $I_i$
according to the permutation $\pi,$
in such a way that the interval in the
$i^{th}$ position, $I_i,$ is translated to the $\pi(i)^{th}$ position
(from left to right). In this way the transformation obtained is of the
form
\[ E(x)=x+d_i,\quad x\in I_i,\, \quad i=1,2,\dots,n,
\] for some displacements $d_1,\dots,d_n.$

In this work we consider mostly iets of the form
$E:[0,b)\to [0,b), b>0$.
An iet $E$ is {\it normalized} if it is defined in the interval
$[0,1)$. The set of all  normalized iet is parameterized in the
following way.
Let $\Delta^{n-1}$ denote the standard simplex of $\R^n,$
\[ \Delta^{n-1} := \left\{(\lambda_1,\lambda_2,\dots,\lambda_n)\in
\R^n_+:
\lambda_1+\lambda_2+\dots+\lambda_n=1\right\}.
\] If $P_n$ is the set of all permutations of the symbols
$\{1,2,\dots,n\},$ then it is usual to identify the product
${P_n}\times \Delta^{n-1}$ with the set of all normalized interval
exchanges of
$n$ intervals. Let us denote by $G_n$  the
set of {\it irreducible} permutations in $P_n,$  i.e., those
permutations
$\pi$ for which
$\pi(\{1,2,\dots,k\})\neq \{1,2,\dots,k\}$ unless $k=n.$
For a fixed permutation $\pi\in G_n$ we
denote by $\ep$ the set of all normalized iets with permutation $\pi$.
We identify the metric spaces $\Delta^{n-1}$ and $\ep$ by the
homeomorphism
$\Delta^{n-1} \ni {\lambda} \mapsto E_{\lambda}:= (\pi,\lambda)$.

Let $\Sigma_n:= \{1,2,\dots,n\}^\Z$. Associated to the orbit of $x$ by
$E_\lambda$ there is the so-called {\it itinerary of $x$},
$\omega_\lambda(x),$ which is an element of $\Sigma_n$ and which is given
by
the natural encoding of the
$E_\lambda$-orbit of~$x$ by assigning to each entry of this orbit the
suffix $\, i\,$ of the interval $I_i$ which contains it.
Set
\[\Omega_{\lambda}={\rm closure}\,\left\{\omega_\lambda([0,1))\right\}
\] in~$\Sigma_n$, so that $\Omega_{\lambda}$ with
the left shift dynamics is a subshift over the alphabet
$\{1,2,\dots,n\}$.

In the next theorem we will consider the $(n-1)$-dimensional  Lebesgue
measure in $\Delta^{n-1} \cong \ep, \pi\in G_n.$
We also recall that a nonempty set in a metric space  is a {\it Cantor
set} if it is closed with empty interior
and no isolated points.

Let us fix an irreducible permutation $\pi\in G_n$ and an arbitrary
injective map~$V:\nn \to \R$. For any $\omega\in \Sigma_n$ consider the
potential
$V(\omega):=(V(\omega_j))_{j\in\Z}$ and the operator $H_{V(\omega)}$ as
in~(\ref{hamiltonian}).

\begin{Theorem}\label{mainThm} There is a subset $\mathcal
F\subset\ep$ of full Lebesgue measure such that, for each $E_{\lambda}\in
\mathcal F,$ the spectrum of the corresponding
Schr\"odinger
operator~(\ref{hamiltonian}) with potential $V(\omega_{\lambda}(x))$
is a Cantor set of zero (Lebesgue) measure and pure  singular continuous
for a.e.~$x\in[0,1)$.
\end{Theorem}

We recall that for $n=2$ and~$\pi(1,2)=(2,1)$ there is only one
discontinuity point~$a_1\in[0,1)$ and the
system is reduced to rotations of the  circle by the angle~$(1-a_1)$.
In
this case the
potentials~$\Omega_{E}$ are the Sturmian  sequences~\cite{BIST,DL1},
which
include the well-known Fibonacci
substitution  sequence~\cite{Q,Su1}. Hence, the potentials generated by
iets are natural generalizations of
Sturmian potentials,  one of the standard models of one-dimensional
quasicrystals. We refer the reader to
\cite{deOG} for additional comments and to
\cite{DLbor}, as well as references therein, for related examples of
Cantor zero measure spectrum.

We close this section with an open question we have found
interesting.
Although almost every iets  are
minimal and uniquely ergodic \cite{Masur,Veech}, there are cases of
minimal iets with more than one ergodic
component ($n/2$ is an upper bound for the number of ergodic
probability
measures \cite{KH,Masur}), so it
is natural to ask if the characterization of the spectrum in
Theorem~\ref{mainThm} holds in such cases.

  \section{Some related results}
The proof of Theorem~\ref{mainThm} will
be reduced to the proof that,
given an irreducible permutation $\pi\in G_n$, for almost every iet
$E_\lambda\in \ep,$ the
corresponding Schr\"odinger operators
$H_{V(\omega_\lambda(x))}$ have no eigenvalues for Lebesgue almost every
$x\in
[0,1)$. In this section we clarify such
statement and the next sections are devoted to the related proof of
absence
of eigenvalues.

 Consider a finite alphabet $\mathcal A$ and let $W$ be the set of
finite words in this alphabet.
If $B\in W$ we denote by $|B|$ its length. Consider a subshift $\Omega$
over $\mathcal A$ (the
dynamics is always given by the left
shift) and, for a given word $B\in W,$ let us denote by
\[
 V_B = V_B(\Omega) = \{\omega\in\Omega: \omega(1)\cdots\omega(|B|)=B
\},
\]the cylinders.
  For each invariant probability measure $\mu$ on $\Omega,$ set
\[
\eta_\mu(n) = \min\{\mu(V_B): B\in W,|B|=n  \}.
\]
In \cite{Bor1,Bor2} Boshernitzan introduced the following condition
(which was later called {\it condition (B)} in \cite{DLbor}):
the subshift $\Omega$ satisfies condition (B) if there
exists
an ergodic probability measure
$\mu$ on
$\Omega$ with
\[
\limsup_{n\to\infty} n\,\eta_\mu(n)>0 \quad {\rm (B)}.  \]
This condition was shown by Boshernitzan \cite{Bor3} to imply unique
ergodicity for minimal subshifts
and, in the particular case of iets, was previously done by Veech
\cite{Veech2}.
Boshernitzan \cite{Bor3} proved
  \begin{Theorem}\label{teorBor3} Let $\pi\in G_n$. Then for Lebesgue
almost every ${\lambda}\in
\Delta^{n-1}$ the subshift $\Omega_{\lambda}$ satisfies condition (B).
\end{Theorem}
\noindent Then, it is obtained  Veech \cite{Veech}
and Masur \cite{Masur} result, that is,
$E_\lambda$ is uniquely ergodic (and, in particular, minimal) for a.e.\
${\lambda}\in
\Delta^{n-1}.$

Recall also the following basic result:
  \begin{Lemma}\cite{Keane1}\label{shiftLemma}  If $E_\lambda$ is
minimal, then $\Omega_{\lambda}$ is  a
minimal subshift, i.e., every $\omega\in\Omega_{\lambda}$ has dense
orbit in~$\Omega_{\lambda}$.
\end{Lemma}

With respect to the spectrum of discrete Schr\"odinger operators, the
following important result was proved in
\cite{DLbor}:

\begin{Theorem}\label{teorDLbor} Let $\Omega$ be a minimal subshift
which satisfies condition (B). If
$\Omega$ is aperiodic, then there exists a Cantor set $\Sigma\subset\R$
of
Lebesgue measure zero so that
the spectrum
$\sigma(H_\omega)=\Sigma$ for every $\omega\in\Omega$.
\end{Theorem}

It is well known that minimality implies the
spectrum (as a set) is the same for all elements in
the hull $\Omega$; by combining this with
Theorem~\ref{teorDLbor} and the
constancy of the absolutely continuous spectrum \cite{LS},  for a fixed
injective map
$V:\{1,2,\dots,n\}\to\R$ we summarize some important known results in

  \begin{Corollary}\label{corolZero} Let $\pi\in G_n.$ There is a  subset
$\mathcal L\subset\ep$ of full Lebesgue measure so that:
\begin{enumerate}
\item[(i)] for each $E_{\lambda}\in \mathcal L$ the spectrum
of~$H_{V(\omega)}$ in~(\ref{hamiltonian}) is
the same  for all~$\omega \in \Omega_{\lambda},$  and it is a Cantor
set of zero Lebesgue measure.
\item[(ii)] for each $E_{\lambda}\in \mathcal L$ the corresponding
Schr\"odinger
operators~(\ref{hamiltonian}) with potentials $V(\omega_\lambda(x))$ have
no
absolutely continuous spectrum
for all $x\in[0,1)$.
\end{enumerate}
\end{Corollary}
Therefore in order to prove Theorem~\ref{mainThm} it is
enough to show that, given $\pi\in
G_n$, there is a set $\mathcal P\subset\ep $ of full Lebesgue measure
so that for each $E_{\lambda}\in
\mathcal P$ the corresponding Schr\"odinger operators
$H_{V(\omega_\lambda(x))}$ have no eigenvalues for Lebesgue almost every
$x\in [0,1)$; thus only singular continuous spectrum remains. Then the
set $\mathcal F$ in Theorem~\ref{mainThm} can be defined by the
intersection of this set $\mathcal P$ with $\mathcal L$.

  An important tool to exclude eigenvalues for a given operator
$H_\omega$, $\omega\in\Sigma_n$, is the Delyon-Petritis \cite{DP1}
version of an argument of
Gordon~\cite{Go}, by means of suitable local word repetitions.

\begin{Theorem}\cite{DP1}\label{thmDP} If for given~$\omega \in
\Sigma_n$ there exists a sequence
$k_i \to
\infty$ such that
\[ {\omega}_{j-k_i}={\omega}_j={\omega}_{j+k_i},
\] for all $1 \le j \le k_i,$ then the Schr\"odinger
operator~$H_{\omega}$ in~(\ref{hamiltonian}) has no
eigenvalues.
\end{Theorem}

Given an irreducible permutation~$\pi$, the idea is to  show that, for
almost all
$\lambda\in\Delta^{n-1}$, Theorem~\ref{thmDP} applies to~$H_\omega$,
$\omega=\omega_\lambda(x)$, with~$x$ in a set of total Lebesgue measure
over~$[0,1)$. In other words, we
have to  prove that for almost all $x\in [0,1)$ there is a sequence
of natural numbers $r_k\to \infty$  such that
{\it the itinerary of $x$ associated to
$[-r_k, 2r_k]$} (i.e., the itinerary of the finite orbit
$\, E_\lambda^{-r_k}(x), E_\lambda^{-r_k+1}(x), \cdots,
E_\lambda^{2r_k}(x)$)
is of the form
\[ \overbrace{\omega_0\omega_1\dots \omega_{r_k}}^{} \,\,\,
\overbrace{\omega_0\omega_1\dots \omega_{r_k}}^{} \,\,\,
\overbrace{\omega_0\omega_1\dots \omega_{r_k}}^{} ,\quad
\omega_i\in\{1,2,\dots,n\}. \]
To prove this statement (which is Proposition~\ref{NoPoint} of
Section~\ref{proofmainthm}) we will use the well-known {\it Rauzy
Renormalization Operator} in the space of
interval exchange transformations. In the next section we will
introduce
the appropriate definitions and
necessary results.

\section{Rauzy's Renormalization} Let $E:[a,b)\to [a,b)$ be a minimal
iet such that $E=(\lambda,\pi)$ for some $\pi\in G_n$ and
$\lambda\in\R^n_{+}.$ Let $J:=[c,d)$ be a proper subinterval of
$[a,b).$ Let us denote by $E_J$ the {\it Poincar\'{e}'s first return
map} of $E$ to the interval $J,$ that is, for $x\in J$, $E_J(x)$ is
given by the  first point in the positive orbit of $x$ (by $E$) that
steps into the interval $J$. By the minimality of $E,$ the map $E_J$
is again an iet of $p$ intervals with $p\ge n$. We will be
interested only in the case $p=n;$ in this way $E_J$ will be
associated to a pair $(\lambda^\prime,\pi^\prime)$ with
$\lambda^\prime\in \R^n_+$ and $\pi^\prime\in G_n$. The iet $E_J$
will be called the {\it induced map of $E$ onto the interval $J.$}
Let $I_1,I_2,\dots, I_n$ be the intervals of continuity of $E_J.$
For each $1\le k\le n$ there exists an integer $r_k > 0$ such that
\[ E(I_k), E^2(I_k), \dots, E^{r_k}(I_k) \] are all intervals disjoints
of
$J$ whereas $E^{r_k+1}(I_k)$ is totally contained in $J.$ By definition
$E_J(I_k) =   E^{r_k+1}(I_k)$. The
number $r_k$ is called the {\it return time of $I_k$ to $J$}.  We
remark that the minimality
of $E$ also implies that the interval $[a,b)$ is given by the union
\begin{equation}\label{equationunionintervals} [a,b) = \bigcup_{k=1}^n
\bigcup_{j=1}^{r_k} E^j (I_k).
\end{equation}

\noindent{\bf Rauzy's map.}  Take $(\pi,\lambda)\in {G_n}\times
\Delta^{n-1}$ and let $E$ be the
$(\pi,\lambda)$-interval exchange.  Consider $\nu=\nu(\pi,\lambda)$
defined as the minimum between
$\lambda_n$ and  $\lambda_{\pi^{-1}(n)}$ provided that these numbers
are
different. If $E_J$ is the
induced map of $E$ onto the interval $J:=[0,1-\nu),\, \nu >0,$ it is
proved
in \cite{Rauzy,Rauzy2} that
$E_J$ is an iet of exactly $n$ intervals associated to a new
irreducible
permutation
$\pi^\prime\in G_n$ and some vector
in $\R^n_+.$ Then, by
normalizing
$E_J,$ we obtain a pair $(\pi^\prime,\lambda^\prime)\in {G_n}\times
\Delta^{n-1}$.
Rauzy's Renormalization map is the association
$(\pi,\lambda)\stackrel{R}{\rightarrow}
(\pi^\prime,\lambda^\prime).$ We remark that
such association
 is not defined when
$\lambda_n = \lambda_{\pi^{-1}(n)}.$ As a consequence of this,
when all iterates $R^n$ of the Rauzy map $R$ are defined
in an iet $E_\lambda = (\pi,\lambda)$, then the orbits of any
pair of discontinuity points of $E_\lambda$ are disjoint and so,
by a Keane's result \cite{Keane1}, $E_\lambda$ is minimal.

It is proved in \cite{Rauzy} that
${G}_n$ is divided into several subsets called {\it Rauzy Classes}
which
are invariant by the process  of
induction just defined. Let us denote by
${\mathcal C}$ one of the  Rauzy classes of $G_n.$ Then if $\pi\in
{\mathcal C}$ and
$R(\pi,\lambda)=(\pi^\prime,\lambda^\prime)$ then $\pi^\prime\in
{\mathcal C}.$  Set
\[
\Delta^{n-1}_{\mathcal C}:=
\left\{(\lambda_1,\lambda_2,\dots,\lambda_n)\in
\Delta^{n-1}: \lambda_{n} \neq
\lambda_{\pi^{-1}(n)},\, \pi\in {\mathcal C} \right\}.
\] Then the transformation
$R(\pi,\lambda)=(\pi^\prime,\lambda^\prime)$ is well defined in
${\mathcal C}\times \Delta^{n-1}_{\mathcal C}.$  In ${\mathcal C}\times
\Delta^{n-1}$ there is a natural measure $m$ which is the product of
the
counting measure in ${\mathcal  C}$
and the $(n-1)$-dimensional Lebesgue measure in $\Delta^{n-1}$. As the
set
$\Delta^{n-1}_{\mathcal  C}$
has total measure in
$\Delta^{n-1}$ (regarding the $(n-1)$-dimensional Lebesgue measure),
the
map $R$ is defined $m$-almost
everywhere in
${\mathcal C}\times \Delta^{n-1}$ and by abuse of language it is
usually
written
$${R}: {\mathcal C}\times \Delta^{n-1}\to  {\mathcal C}\times
\Delta^{n-1}.$$
  The Rauzy's transformation plays  a central role in the ergodic
theory
of iets due to the following
result \cite{Veech, Masur}.
  \begin{Theorem} The Rauzy's operator is ergodic for a
measure which is absolutely continuous with respect to the
probability measure $m$ defined above.
\end{Theorem}
   Another related result proved in \cite{Veech, Masur} is
  \begin{Theorem}\label{corollaryalmostalllambda} For a fixed
permutation
$\pi\in  {\mathcal C}$ and for almost every vector
$\lambda\in\Delta^{n-1},$ the $R$-orbit of
$E_\lambda$ is dense in ${\mathcal C }\times
\Delta^{n-1}$. \end{Theorem}

\section{Proof of Theorem~\ref{mainThm}}\label{proofmainthm}

First at all, let us show a relation between the process of
renormalization and
the key property in Theorem~\ref{thmDP}. Let $E\colon [0,1)\to[0,1)$ be
a minimal interval
exchange and $E_I$ the induced map
of $E$ onto $I=[a,b)\subset [0,1).$ Suppose that $L\subset I$ is an
interval of continuity of  $E_I$ and
that, for some $x\in L,$
$\{E_I^{-1}(x), E_I(x)\}\subset L.$ If $r$
is the return time of $L$
to $I$ and $B:=\omega_0 \omega_1\dots \omega_r,\, \omega_j\in\{1,2,\dots,n\},$ is the
itinerary of $x$ until it returns to $L,$
(i.e, $E^j(x)$ is in the $\omega_j$-interval
of continuity of $E$ for each $\, 0\le j\le r$),
then the itinerary of every point in $L,$ of  length $r$,
is given by the same word  $B.$
Therefore as
$E_I(x)\in L$ and $E_I^{-1}(x)\in L,$ the itinerary of $x$ associated
to  $[-r,2r]$ will be given by the word
\begin{equation}\label{eqpointrecorrent}
\overbrace{\omega_0\omega_1\dots \omega_r}^{B}\,\,\,
\overbrace{\omega_0\omega_1\dots \omega_r}^{B} \,\,\,
\overbrace{\omega_0\omega_1\dots \omega_r}^{B}.
\end{equation}
One such a point
will be called {\it a candidate point in the interval $[a,b)$ for the length $r$.}

Observe that if $E_I^{-1}(x), x, E_I^{}(x)$ and $E_I^{2}(x)$ belong to $L$,
then the itinerary of each of the points $y=E^k(x), 0\le k\le r$ associated  to
$[-r,2r]$ is of the form $WWW$ where $W$ is the word
\[ \omega_k\omega_{k+1}\dots \omega_r\omega_0\omega_1\dots \omega_{k-1}, \]
this is, $y$ is also a candidate point for the length $r$. Therefore
we have

\begin{Lemma}\label{lemma-candidate-points}
Under the assumption right above, assume that
\[ x\in E_I^{-2}(L)\cap E_I^{-1}(L)\cap L \cap E_I(L).\] Then
$x,E(x),E^2(x),\dots,E^{r}(x)$ are
all candidate points for the length~$r$.
\end{Lemma}

Note that if there is a nested sequence of intervals shrinking to  a
point
$x$,
\[ [a_1,b_1) \supset [a_2,b_2) \supset \dots [a_k,b_k) \supset\dots \]
and
 such that $x$ is a candidate
point for each $[a_k,b_k)$ for the length $r_k,$ then necessarily
$r_k\to\infty$ as
$k\to\infty$, and the itinerary of $x$ associated to its
whole orbit is given by a sequence that satisfies the hypotheses of
Theorem~\ref{thmDP}.

\medskip

\noindent{\bf The periodic iet $P$.}\label{keyexample} Let us
consider the following family of periodic iets. Let $n\in \N$ be a
natural number and let $\pi\in G_n$ be an arbitrary irreducible
permutation. Consider the vector
\[ \lambda^*:=(\frac{1}{n}, \frac{1}{n}, \dots, \frac{1}{n})\in \R^n; \]
then, the intervals of continuity of the iet $P:=(\pi, \lambda^*)$
are of the form
\[ I^*_k:=[\frac{k}{n},\frac{k+1}{n}),\, 0\le k \le n-1,\] and
$P$ sends any interval $I^*_k$  {\it onto} the interval
$I^*_{\pi^{}(k)}.$ In this way, for each $k\in \{ 1,2,\dots,n\},$
there is a positive integer $l_k$ such that
 \[ P^{l_k+1} (I^*_k) =  I^*_k; \]
in particular, $P$ is a periodic iet; that is, for some $N\in \N,$
$P^N$ is the identity map. We may suppose that $l_k$ is the smallest
positive integer with this property. Every $x \in I^*_k$ is a
periodic point of period $l_k$ and therefore a candidate point for
the length $l_k$. The induced map of $P$ in $I^*_k$ is the identity.
If $E$ is an iet very close to $P$ and $I$ is an interval of
continuity of $E$, the induced map $E_I$ in $I$ is very ``close to
the identity:" there is an interval of continuity of $E_I$ whose
length is very close to that of $I.$ By shortening a little this
interval we obtain an interval $L$ such that the length of
$E_I^{-1}(L)\cap E_I^{-1}(L)\cap L \cap E_I^{}(L)$ is still close to
the length of $I.$ Lemma~\ref{lemma-candidate-points} guaranties
that all the points in this intersection are candidate points for
the length $l_k.$

By the comment that followed Theorem~\ref{thmDP}, the proof of
Theorem~\ref{mainThm} is a consequence of the following proposition.
   \begin{Proposition}\label{NoPoint} Fix an irreducible permutation
$\pi\in G_n.$ Then for almost all
${\lambda}\in \Delta^{n-1}$ and for almost all $x\in[0,1),$
the coding~$\omega_{\lambda}(x)$ satisfies the hypotheses of
Theorem~\ref{thmDP}.
   \end{Proposition}

\begin{proof}
Consider the periodic iet $P=(\pi,\lambda^*)$, where
$\lambda^*:=(\frac{1}{n}, \frac{1}{n}, \dots, \frac{1}{n})$, and let
$l_k, 1\le k \le n-1,$ be the smallest positive integer such that
$P^{l_k}([\frac{k}{n},\frac{k+1}{n}))=[\frac{k}{n},\frac{k+1}{n})$.

For any small number $\delta>0$, let $\vartheta(\delta)$
be the set of iets $E=(\pi,\lambda)$ with
 \[ \max \{|\lambda_i - \frac{1}{n}| : 1 \le i \le n \} < \delta. \]

Fix $0< \epsilon <1.$ Given a natural number $m$, there exists
$\delta_m>0$ such that if $E\in \vartheta(\delta_m)$ and $k\in \{
1,2,\dots,n\},$ then the  $k^{th}$-interval of continuity of $E,$
denoted by $I_k$, contains an interval $L$ such that the sets
$E_I^{-2}(L)$, $E^{-1}(L)$ and $E^{}(L)$ are subintervals of $I_k$
and the interval
\[ M_k:= E_{I_k}^{-2}(L)\cap E_{I_k}^{-1}(L)\cap L \cap E_{I_k}^{}(L)\]
satisfies
 \begin{equation}\label{eq:C/J}
    \frac{|M_k|}{|I_k|} \ge 1 - \frac{\epsilon}{2^m}. \end{equation}
where $|A|$ denotes the Lebesgue measure of $A.$

Recall that the Rauzy orbit of $E_\lambda:=(\pi,\lambda)$ is dense
in ${\mathcal C}\times \Delta^{n-1}$ for almost all
$\lambda\in\Delta^{n-1}$ (Theorem~\ref{corollaryalmostalllambda}).
Take an iet $E$ whose Rauzy orbit is dense. Then there are
\begin{itemize}
\item[a1)] a nested sequence of half-open intervals
\[ J_1 \supseteq J_2 \supseteq \dots \supseteq J_m \dots \] given by
the Rauzy process of induction, \item[a2)] a sequence of linear
bijections $H_m: J_m \to [0,1),$ and \item[a3)] an increasing
sequence $(N_m)_{m\ge 1}$ of natural numbers such that
$E_m:=R^{N_m}(E) = H_m\circ E_{J_m}\circ H_m^{-1}$  belongs to
$\vartheta(\delta_m)$, where $E_{J_m}$ is the map induced by $E$
onto $J_m.$ \end{itemize} Fix $m\in \N$ and, for each $k$ in
$\{1,2,\dots,n\},$ let $I_k$ be the $k^{th}$ interval of continuity
of $E_m$ and $M_k\subset I_k$ be a set that
satisfies~(\ref{eq:C/J}). Let $r_k$ be the return time, with respect
to $E,$ of the interval $I_k$ to $J_m.$ By
Lemma~\ref{lemma-candidate-points}, all the iterates
\[M_k, E(M_k), \dots, E^{r_k}(M_k) \]
are made of candidate points for the length $r_k.$ As $E$ preserves
the Lebesgue measure, each of these sets has Lebesgue measure
greater than $1-\epsilon\cdot 2^{-m}$ times the  measure of $I_k$.
This implies that the set of candidate points, for the length $r_k,$
in the union
\[ \bigcup_{j=1}^{r_k} E^j(I_k) \] is greater than $1-\epsilon\cdot
2^{-m}$ times the Lebesgue measure of this
union.

Let $C_m$ be the set of candidate points in $[0,1)$ for lengths $r \le
r_m:=\max\{r_1,\dots,r_n\}.$ Using the fact that
\[ [0,1) = \bigcup_{k=1}^n \bigcup_{j=1}^{r_k} E^j(I_k), \] (see
relation~(\ref{equationunionintervals})) we conclude that
$|C_{m}|  \ge 1 -\frac{\epsilon}{2^m}.$

Now it follows easily that the Lebesgue measure of the
intersection
\[ C_\epsilon := \bigcap_{m \ge 1} C_{m} \] is greater than $
1 - \epsilon \,\sum_{m \ge 1} 2^{-m}=1- \epsilon.$ Observe that
$(r_m)$ is a strictly increasing sequence and that, for all
$m=1,2,\cdots,$ every point in $C_\epsilon$ is a candidate point for
the length $r_m.$ Therefore all the points in $C_\epsilon$ satisfy
the condition of Proposition~\ref{NoPoint}. As the number
$\epsilon>0$ is arbitrary, Proposition~\ref{NoPoint} is proved.
\end{proof}

\subsubsection*{Acknowledgments} {\small CRdeO was partially
supported by CNPq.  CG thanks the partial support by FAPESP
Grant 03/03107-9 and by CNPq Grant 306992/2003-5, Brazil.
MC was partially supported by FAPES grant \\ 30898951/2005.}

\end{document}